\documentclass[sigconf]{acmart}
\AtBeginDocument{%
  }

\copyrightyear{2026}
\acmYear{2026}
\setcopyright{cc}
\setcctype{by}
\acmConference[CIKM '26]{Proceedings of the 35th ACM International Conference on Information and Knowledge Management}{November 07--11, 2026}{Rome, Italy}
\acmBooktitle{Proceedings of the 35th ACM International Conference on Information and Knowledge Management (CIKM '26), November 07--11, 2026, Rome, Italy}
\acmDOI{10.1145/3799682.3840731}
\acmISBN{979-8-4007-2539-5/2026/11}

\usepackage{xspace}
\usepackage{multirow}
\usepackage[capitalise]{cleveref}
\usepackage[table]{xcolor}
\usepackage{colortbl}
\usepackage{subcaption}
\usepackage{balance}
\usepackage{tcolorbox}
\usepackage{orcidlink}
\usepackage{twemojis}

\makeatletter
\newcommand{\emoji}[1][1em]{\@emojiaux{#1}}
\newcommand{\@emojiaux}[2]{%
  \raisebox{-0.22\height}{\resizebox{!}{#1}{\@nameuse{emoji@#2}}}%
}
\@namedef{emoji@woman-surfing}{\twemoji{woman surfing}}
\@namedef{emoji@red-question-mark}{\twemoji{red question mark}}
\@namedef{emoji@check-mark-button}{\twemoji{check mark button}}
\makeatother

\newcommand{\up}{$\uparrow$\xspace}
\newcommand{\down}{$\downarrow$\xspace}

\newcommand{\second}[1]{\underline{#1}}
\newcommand{\approach}{SURF\xspace}
\newcommand{\ndatasets}{7\xspace}
\definecolor{lightgray}{gray}{0.92}

\newcommand{\red}[1]{\textcolor{black}{#1}}

\begin{document}

\title{\texorpdfstring{\emoji{woman-surfing}}{} \approach: Subtractive Updates for Recommender Forgetting}

\author{Filippo Betello\textsuperscript{\textdagger} \orcidlink{0009-0006-0945-9688}}
\affiliation{%
  \institution{Sapienza University of Rome}
  \city{Rome}
  \country{Italy}}
\email{betello@diag.uniroma1.it}
\authornote{\textdagger Equal Contribution}

\author{Antonio Purificato\textsuperscript{\textdagger} \orcidlink{0009-0009-3933-380X}}
\affiliation{%
  \institution{Sapienza University of Rome}
  \city{Rome}
  \country{Italy}}
\email{purificato@diag.uniroma1.it}

\author{Nicola Tonellotto \orcidlink{0000-0002-7427-1001}}
\affiliation{%
  \institution{University of Pisa}
  \city{Pisa}
  \country{Italy}}
\email{nicola.tonellotto@unipi.it}

\author{Fabrizio Silvestri \orcidlink{0000-0001-7669-9055}}
\affiliation{%
  \institution{Sapienza University of Rome}
  \city{Rome}
  \country{Italy}}
\email{fsilvestri@diag.uniroma1.it}

\renewcommand{\shortauthors}{Filippo Betello, Antonio Purificato, Nicola Tonellotto, \& Fabrizio Silvestri}


\begin{abstract}
The increasing demand for user privacy and compliance with regulations such as GDPR has made machine unlearning a fundamental requirement for modern recommender systems. However, Sequential Recommender Systems (SRS) pose unique challenges for unlearning due to their reliance on temporal interaction patterns. Existing approaches either require computationally prohibitive full retraining or fail to account for the sequential nature of user behavior.
We propose \approach (\textbf{S}ubtractive \textbf{U}pdates for \textbf{R}ecommender \textbf{F}orgetting), a lightweight framework for approximate machine unlearning in SRS. \approach operates in three stages: (i) identifying the neighborhood of the item to forget in the embedding space, (ii) training an auxiliary model on this compact local subset, and (iii) subtracting the auxiliary model's scores from the original model at inference time.
Experiments against five baselines on \ndatasets\ datasets show that \approach achieves unlearning effectiveness comparable to full retraining while substantially reducing computational cost, yielding up to a 32\% improvement in NDCG\red{@20} \red{while requiring just 2\% of the original retraining baseline time budget}. We share our code at \url{https://github.com/FilippoBetello/SURF}.
\end{abstract}

\ccsdesc[500]{Information systems~Recommender systems}
\ccsdesc[300]{Computing methodologies~Neural networks}
\ccsdesc[100]{Social and professional topics~Copyrights}

\keywords{Recommender Systems, Sequential Recommendation, Machine Un-learning, Recommendation Unlearning, Approximate Unlearning, Influence Subtraction.}

\maketitle

\begin{figure*}
    \centering
    \includegraphics[width=\linewidth]{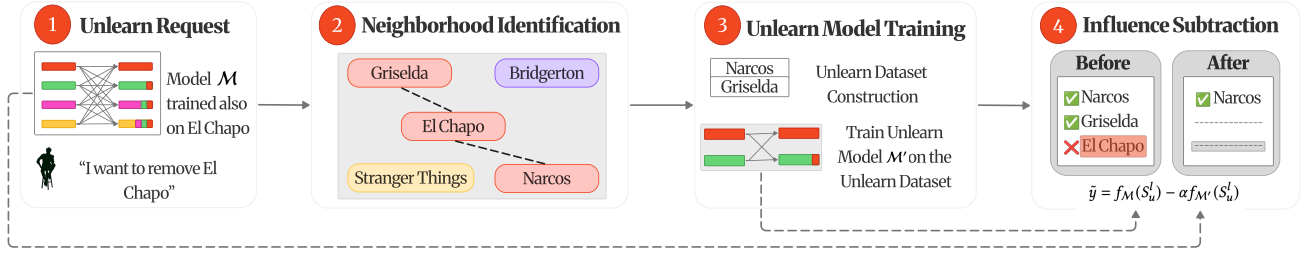}
    \caption{\looseness -1 Overview of \approach: Given an unlearning request, we identify the neighborhood of the item to forget, train an auxiliary model on this subset, and subtract its influence at inference time removing the target item from recommendations.}
    \label{fig:graphical_abstract}
\end{figure*}

\section{Introduction}
Recommender systems play a pivotal role in today's digital landscape, influencing user experiences in e-commerce, entertainment, and social media \cite{purificato2025sheaf4rec,10128133}. Among these, Sequential Recommender Systems (SRSs) exploit temporal and behavioral signals to model how user preferences evolve over time. Through the analysis of interaction sequences, SRSs are able to provide highly personalized and contextually relevant recommendations \cite{pan2026survey,siciliano2025convolutional}. However, their data-intensive nature also raises significant concerns regarding privacy, data ownership, and regulatory compliance~\cite{10.1145/3460231.3473326}.

In recent years, growing attention to data protection laws, such as the General Data Protection Regulation (GDPR) \cite{voigt2017eu} in Europe and similar frameworks worldwide \red{\cite{pardau2018california}}, has introduced the ``right to be forgotten''\footnote{\url{https://gdpr.eu/right-to-be-forgotten/}}, granting individuals the ability to request the deletion of their personal data. This has given rise to an emerging research area known as machine unlearning, which seeks to develop algorithms capable of efficiently removing the influence of specific data points\red{, such as images in classification tasks or items in recommendation scenarios}, from trained models, without requiring full retraining from scratch \cite{nguyen2025survey}. \red{Early machine unlearning approaches primarily focused on supervised learning models, aiming to efficiently remove subsets of training data without retraining from scratch \cite{bourtoule2021machine}.}
Although \red{machine unlearning} has been thoroughly investigated in the context of classical matrix factorization and GNN-based recommender systems \cite{receraser,chen2025pre}, SRSs pose unique difficulties for unlearning: these models rely on historical user sequences and temporal dependencies, removing items can affect multiple learned representations~\cite{wang2019sequential}.

Existing recommendation unlearning methods mainly fall into two categories: partition-based \cite{receraser,ultrare,sru} and influence-function-based approaches \cite{wu2023gif,ifru}. Partition-based methods divide training data into shards and retrain only affected shards upon unlearning requests; however, in recommendation systems this strategy is ineffective because users and items are highly interconnected \cite{singh2020scalability}, making items likely to appear across many shards and degrading recommendation accuracy while increasing retraining cost. Influence-function-based methods estimate the impact of data removal without retraining, but they rely on strong mathematical assumptions that often break in practice~\cite{basu2021influence} and incur high memory overhead~\cite{zhu2025revisiting}. These limitations motivate the need for more efficient and reliable unlearning solutions.

To address these limitations, we propose \textbf{\approach} (\textbf{S}ubtractive \textbf{U}pdates for \textbf{R}ecommender \textbf{F}orgetting), a novel unlearning paradigm bridging partition-based and influence-function-based approaches. \red{By combining the strengths of both paradigms, we bridge rigorous theoretical guarantees, supported by formal mathematical analysis, with strong empirical validation, while avoiding data sharding, which is unsuitable in SRSs due to the dense and highly interdependent structure of user–item interactions.}
Given an item to be forgotten, we first identify the subset of users who have interacted with it. We then train a lightweight auxiliary model on this restricted set of interactions, so that the model is trained exclusively on the data to be forgotten. To the best of our knowledge, this is the first approach introducing model subtraction as an unlearning mechanism in recommender systems.
During inference, unlearning is achieved by subtracting the scores of the auxiliary model from those of the original model, effectively removing the influence associated with the forgotten item. 
Unlike post-hoc filtering methods, which operate solely at the output level and leave the model’s internal representations unchanged, \approach affects the prediction process at the representation level. Specifically, the method subtracts the contribution of an auxiliary unlearn model trained on a localized neighborhood of the forgotten item, thereby attenuating its influence on both latent representations and downstream sequential predictions. This approach is particularly well-suited for sequential recommendation, where item removal has cascading effects throughout user sequences \cite{pan2026survey}. \red{An additional advantage of \approach is that the forgetting process is inherently reversible. Contrary to parameter-editing or retraining-based approaches, \approach preserves the original model unchanged. Consequently, reverse unlearning can be performed by simply restoring the original inference function without requiring additional optimization, making it suitable for real-world production.}
The proposed strategy is model-agnostic and enables efficient unlearning across \red{State Of The Art (SOTA)} recommender architectures, achieving effective forgetting with minimal computational overhead. While \red{our method }relies on an approximate, \red{but theoretically grounded,} isolation of item influence, empirical results show consistent improvements of up to 32\% \red{in NDCG@20} over existing methods, \red{while requiring just 2\% of the original retraining baseline time budget,} highlighting a favorable trade-off between unlearning effectiveness and efficiency. Furthermore, we provide theoretical results formally justifying the proposed mechanism, showing that \approach is theoretically grounded.

\section{Related Works}
\label{sec:related_work}
\red{In this section, we review existing research on SRSs and recommendation unlearning methods, and conclude with our contributions and distinguishing factors from previous work.}

\subsection{Sequential Recommendation}

Sequential recommendation emerged from Markov chain–based approaches~\cite{fouss2005web}, which offered strong computational efficiency but were fundamentally limited in their ability to model long-range dependencies in user behavior.
With the rise of deep learning, this paradigm shifted toward Recurrent Neural Networks (RNNs)~\cite{donkers2017sequential}, which represent user interaction histories through continuously evolving latent states. Notable models such as GRU4Rec~\cite{hidasi2015session} leverage Gated Recurrent Units to capture temporal dynamics in sequential data. Despite their success, RNN-based methods remain challenged by difficulties in retaining long-term dependencies and in promoting recommendation diversity.
These limitations have motivated a transition toward attention-based architectures~\cite{vaswani2017attention}. By selectively weighting past interactions, Transformer-based models, including SASRec~\cite{kang2018selfattentive} and BERT4Rec~\cite{sun2019bert4rec}, have achieved state-of-the-art performance in sequential recommendation tasks. Due to their strong empirical performance and extensive use in prior work~\cite{pan2026survey}, GRU4Rec, SASRec, and BERT4Rec have become standard benchmarks in sequential recommendation. Accordingly, we adopt these models as representative baselines in our study.

\subsection{Recommendation Unlearning} \label{sec:related_unlearning}
Machine \red{u}nlearning \cite{nguyen2025survey} has gained increasing attention due to regulatory requirements such as data protection laws and ``the right to be forgotten''~\cite{voigt2017eu}. Early machine unlearning \red{methods} primarily focused on supervised learning models, aiming to efficiently remove subsets of training data without retraining from scratch \red{\cite{bourtoule2021machine}}. 
\red{More recently, unlearning has been explored in generative LLMs through fine-tuning with alternative labels \cite{eldan2023whosharrypotterapproximate}, and in multi-task vision-language models via task arithmetic, where negating a task vector in weight space degrades performance on a target task \cite{ilharcoediting}. While the latter shares with our work the intuition of subtracting learned information, it operates by permanently modifying model weights along task-specific directions and requires a full fine-tuning step to construct each task vector. Our approach instead performs subtraction at inference time on prediction scores, preserving the original model intact, and leverages the local geometry of the embedding space to precisely target the item to be forgotten.}

Recently, unlearning has been explored in the context of recommendation \cite{chen2024cure4rec,yang2026curriculumapproximateunlearningsessionbased}. RecEraser \cite{receraser} is among the first works to adapt SISA to the task of recommendation by sharding the training data, while UltraRE \cite{ultrare} further improves this paradigm through a different shard construction strategy. Other approaches have focused on specific model classes, such as kNN \cite{schelter2023forget}, whereas SRU \cite{sru} extends shard-based unlearning to session-based recommendation.

In parallel, influence function–based methods have been proposed to approximate the effect of data removal without full retraining, including SCIF \cite{li2023selective}, IFRU \cite{ifru}, and UnlearnRec \cite{chen2025pre}. Finally, recent studies emphasize the challenge of evaluating unlearning effectiveness, as existing metrics (e.g., MIA-based) are often difficult to interpret and use in practice \cite{10.1145/3705328.3748092}.

\red{Overall, existing partition-based methods suffer from the need for costly pre-sharding and repeated shard-level retraining, while influence-function-based approaches rely on expensive and often unstable second-order approximations such as Hessian computations. In contrast, \approach bridges these two paradigms by identifying embedding neighborhoods associated with the target item and isolating its local influence without requiring any data partitioning or second-order optimization. This localized influence is then captured via a lightweight auxiliary model trained only on the affected subset, and unlearning is achieved by subtracting its predictions from the original model at inference time. This design enables model-agnostic unlearning for sequential recommendation while avoiding the computational and structural limitations of prior approaches.}

\section{Method}
\red{In this section, we present \approach and describe the overall procedure, including neighborhood identification, construction of the unlearning dataset, auxiliary model training, and influence subtraction. We then provide a theoretical analysis by investigating three key properties: (i) ranking stability, showing under which conditions the relative ordering of relevant items is preserved after unlearning; (ii) local perturbation, quantifying how much the original prediction function is modified within the neighborhood of the forgotten item; and (iii) influence removal, explaining why subtracting the auxiliary model effectively suppresses the contribution of the target item while preserving the remaining predictive signal.
}

\subsection{\approach}
Let $\mathcal{U} \subset \mathbb{N^+}$ be a set of users and $\mathcal{I} \subset \mathbb{N^+}$ a set of items. Each user $u \in \mathcal{U}$ is represented by a sequence of $\ell_u$ interactions $S_u = (i_1, i_2, \ldots,i_{\ell_u})$ with which it has interacted, where $\ell_u$ is the length of the sequence and $i_j \in \mathcal{I}$ is an item. An SRS $\mathcal{M}$ takes as input the user's sequence up to the $l$-th item, $S_u^l = [i_1, \dots, i_l]$, aiming to predict the next item $i_{l+1}$. The $(l+1)$-th item considered the positive item while all the other $l$ items are negative items.

Given a SRS model $\mathcal{M}$ trained on an original dataset $\mathcal{D}$, an unlearning request $\mathcal{R}$ specifies an item $i^* \in \mathcal{I}$ to be forgotten. The goal is to obtain a model $\mathcal{M}'$ that behaves as if it had never been trained on interaction sequences containing $\mathcal{R}$, without retraining $\mathcal{M}$ from scratch.
Let $\mathcal{D}_f \subset \mathcal{D}$ denote the forget dataset, containing all interactions involving the \red{item} to be forgotten:
\begin{equation*}
\mathcal{D}_f = \{S_u : u \in \mathcal{U} \wedge i^* \in S_u\}
\end{equation*}
\Cref{fig:graphical_abstract} provides an overview of the approach, whose steps are detailed below.

\noindent \textit{1. Neighborhood Identification.} Let $\mathbf{e}_{i^*} \in \mathbb{R}^d$ denote the embedding of the item to be forgotten (i.e., El Chapo in \Cref{fig:graphical_abstract}) in the learned embedding space of $\mathcal{M}$. We retrieve its $k$ nearest neighbors:
\begin{equation*}
    \mathcal{N}_k(\mathbf{e}_{i^*}) = \text{$k$-argmin}\sum_{\mathbf{e'} \in \mathcal{E}} \|\mathbf{e}-\mathbf{e'}\|.
\end{equation*}

\red{Where $\mathcal{E}$ is the set of item embeddings in the model, i.e., $\mathcal{E=\{e_j \mid j \in \mathcal{I}\}}$}. For architectures without user embeddings (e.g., SASRec~\cite{kang2018selfattentive}), we represent users by aggregating their item embeddings:
\begin{equation*}
\mathbf{e}_{u} = \frac{1}{\ell_u}\sum_{j=1}^{\ell_u} \mathbf{e}_{i_j}
\end{equation*}
We rely on the pure $k$-nearest neighbors of $\mathbf{e}_i^*$ rather than only on the embeddings of sequences in $D_f$, since the latter do not fully characterize the local region of the representation space where information about the item is encoded. The neighborhood $\mathcal{N}_k(\mathbf{e}_{i^*})$ provides a better approximation of this local manifold, enabling more effective and robust forgetting \cite{koh2017understanding}.

\noindent \textit{2. Unlearn Dataset Construction.} We construct a small unlearn dataset~$\mathcal{D}_r$ based on the identified neighborhoods, selecting the embeddings most affected by the forgotten item:
\begin{equation*}
    \mathcal{D}_r = \{S_u : \exists \, i \in S_u | i \in \mathcal{N}_k(\mathbf{e}_{i^*})\}
\end{equation*}
\red{Where $\mathcal{D}_r$ denotes the unlearning dataset composed of sequences containing items in the neighborhood of the forgotten one.}

\noindent \textit{3. Unlearn Model Training.} We train a small unlearn model $\mathcal{M}'$ on $\mathcal{D}_r$ using the same architecture and training objective as the original model $\mathcal{M}$. This unlearn model learns to approximate the influence of the unlearn dataset on predictions.

\noindent\textit{4. Influence Subtraction. } At inference time, for a given input sequence $S_u^l$, the new prediction scores $\tilde{y} \in \mathbb{R}^{|\mathcal{I}|}$ are computed as:
\begin{equation*}
    \tilde{y} = f_{\mathcal{M}}(S_u^l) - \alpha f_{\mathcal{M}'}(S_u^l),
\end{equation*}
where $f_{\mathcal{M}}(\cdot)$ returns the prediction scores of the model $\mathcal{M}$ over all items, and $\alpha > 0$ is a scaling hyperparameter controlling the unlearning strength. The final recommendation $\tilde{i}_{l+1}$ is obtained by selecting the item corresponding to the greatest prediction score in $\tilde{y}$:
\begin{equation*}
\tilde{i}_{l+1} = \text{argmax}_{i \in \mathcal{I}}\tilde{y}_i
\end{equation*}

\subsection{Ranking Stability}
\label{sec:stability}

\begin{tcolorbox}[
    colback=red!5,
    colframe=red!30,
    left=1mm
]
\begin{tabular}{@{}m{0.02\linewidth}m{0.90\linewidth}@{}}
\centering \emoji{red-question-mark} &
How can we guarantee that the unlearning operation preserves the relative ranking between relevant items while removing the influence of the forgotten item?
\end{tabular}
\end{tcolorbox}

Given two items $i_1 \text{ and } i_2$, we define:
\begin{equation*}
    \Delta_{i_1 i_2} = f_{\mathcal{M}}(S_u^l)_{i_1} - f_{\mathcal{M}}(S_u^l)_{i_2}
\end{equation*}
\textbf{Theorem}: Given two items $(i_1,i_2)$, if $\Delta_{i_1 i_2}>2\alpha ||f_{\mathcal{M'}}(S_u^l)||_{\infty}$ then the relative ranking using \approach is kept.

\noindent \textbf{Proof}: 
\begin{align*}
    & \tilde \Delta_{i_1 i_2} = \tilde{y}_{i_1} - \tilde{y}_{i_2} = \Delta_{i_1 i_2} - \alpha(f_{\mathcal{M'}}(S_u^l)_{i_1} - f_{\mathcal{M'}}(S_u^l)_{i_2}) \\
    & |f_{\mathcal{M'}}(S_u^l)_{i_1} - f_{\mathcal{M'}}(S_u^l)_{i_2}| \leq 2||f_{\mathcal{M'}}(S_u^l)||_{\infty} \text{ (Triangle Inequality)} \\
    & \tilde \Delta_{i_1 i_2} \geq \Delta_{i_1 i_2} - 2\alpha ||f_{\mathcal{M'}}(S_u^l)||_{\infty}
\end{align*}
If $\Delta_{i_1 i_2}(S_u^l) > 2\alpha ||f_{\mathcal{M'}}(S_u^l)||_{\infty}$ then $\tilde \Delta_{i_1 i_2} > 0$ and the order is kept. Also post-unlearning relevant items remain relevant. $\square$

\red{\noindent \textbf{Interpretation:} \approach preserves the relative ranking between items whenever their original score difference is sufficiently larger than the perturbation introduced by the auxiliary model. Consequently, highly relevant items tend to maintain their ordering after unlearning, while the forgetting operation mainly affects predictions associated with the target item's influence.}

\subsection{Local Perturbation}

\begin{tcolorbox}[
    colback=red!5,
    colframe=red!30,
    left=1mm
]
\begin{tabular}{@{}m{0.02\linewidth}m{0.90\linewidth}@{}}
\centering \emoji{red-question-mark} &
To what extent does the proposed unlearning procedure modify the original scoring function inside the neighborhood of the forgotten item?
\end{tabular}
\end{tcolorbox}

\noindent 
\textbf{Assumption}: The auxiliary model $f_{\mathcal{M}'}$ is Lipschitz continuous in the embedding space, i.e., there exists $L>0$ such that for all embeddings $\mathbf e,\mathbf e' \in \mathcal{N}_k(\mathbf e_{i^*})$:
\begin{equation*}
\left\| f_{\mathcal M'}(\mathbf e) - f_{\mathcal M'}(\mathbf e') \right\| \le L \| \mathbf e-\mathbf e'\|
\end{equation*}
\red{This assumption ensures that nearby embeddings lead to similar auxiliary predictions, ensuring smooth local unlearning effects.}

\noindent \textbf{Assumption}: The influence of the forgotten item is bounded:
\begin{equation*}
\| f_{\mathcal M'}(\mathbf e_{i^*}) \| \le C
\end{equation*}
For some constant $C>0$. \red{This assumption ensures that the forgotten item has limited influence, preventing excessively large unlearning corrections.}

\noindent \textbf{Theorem}: Given $\tilde y= f_{\mathcal M}(S_u^l) - \alpha f_{\mathcal M'}(S_u^l)$, for every sequence $S_u^l$ whose representation
$\mathbf e_u \in \mathcal N_k(\mathbf e_{i^*})$, the perturbation induced by \approach satisfies:
\begin{equation*}
\|\tilde y - f_{\mathcal M}(S_u^l) \| \le \alpha \Big( L\,\mathrm{diam}(\mathcal N_k(\mathbf e_{i^*})) + C \Big)
\end{equation*}
Where:
\begin{equation*}
\mathrm{diam}(\mathcal N_k(\mathbf e_{i^*})) = \max_{\mathbf e_1,\mathbf e_2 \in \mathcal N_k(\mathbf e_{i^*})} \| \mathbf e_1-\mathbf e_2 \|
\end{equation*}

\noindent \textbf{Proof}: Consider an input embedding $\mathbf e_u$ in the neighborhood $\mathcal N_k(\mathbf e_{i^*})$, adding and subtracting $f_{\mathcal M'}(\mathbf e_{i^*})$ gives:
\begin{align*}
\| f_{\mathcal M'}(\mathbf e_u) \| 
&= \| f_{\mathcal M'}(\mathbf e_u) - f_{\mathcal M'}(\mathbf e_{i^*}) + f_{\mathcal M'}(\mathbf e_{i^*}) \| \\
&\le \| f_{\mathcal M'}(\mathbf e_u) - f_{\mathcal M'}(\mathbf e_{i^*}) \| + \| f_{\mathcal M'}(\mathbf e_{i^*})
\| \\
& \le L\|\mathbf e_u - \mathbf e_{i^*} \| + C \\
&\le L\, \mathrm{diam} (\mathcal N_k(\mathbf e_{i^*})) +C
\end{align*}
Therefore:
\begin{align*}
\| \tilde y - f_{\mathcal M}(S_u^l) \|
&= \alpha \| f_{\mathcal M'}(S_u^l) \| \\
&\le \alpha \Big( L\,\mathrm{diam} (\mathcal N_k(\mathbf e_{i^*})) + C \Big)
\end{align*}

Hence, the perturbation introduced by the unlearning operation is bounded.
$\square$

\red{\noindent \textbf{Interpretation:} The theorem provides an upper bound on the magnitude of the changes introduced by \approach within the neighborhood of the forgotten item. The perturbation depends on the scaling parameter $\alpha$, the smoothness of the auxiliary model, and the size of the local neighborhood. Therefore, when the neighborhood is compact and $\alpha$ is controlled, \approach introduces only limited modifications to the original prediction function, promoting localized forgetting while preserving global recommendation behavior.}

\subsection{Influence Removal}

\begin{tcolorbox}[
    colback=red!5,
    colframe=red!30,
    left=1mm
]
\begin{tabular}{@{}m{0.02\linewidth}m{0.90\linewidth}@{}}
\centering \emoji{red-question-mark} &
Why does subtracting the auxiliary unlearning model effectively remove the contribution of the target item from the final predictions?
\end{tabular}
\end{tcolorbox}

\noindent \textbf{Assumption}:
Let the prediction function of the original model admit the decomposition:
\begin{equation*}
f_{\mathcal M}(S_u^l) = f_{\mathrm{base}}(S_u^l) + f_{i^*}(S_u^l),
\end{equation*}
where $f_{i^*}(S_u^l)$ denotes the contribution associated with the target item $i^*$ to be forgotten.
Assume that the auxiliary model $\mathcal M'$ trained on
$\mathcal D_r$ approximates such contribution, i.e., there exists $\varepsilon>0$ such that:
\begin{equation*}
\left| f_{\mathcal M'}(S_u^l) - f_{i^*}(S_u^l)
\right| \le \varepsilon
\end{equation*}

\noindent \textbf{Theorem}:
Given the influence subtraction mechanism $\tilde y = f_{\mathcal M}(S_u^l) - \alpha f_{\mathcal M'}(S_u^l)$,
the residual contribution associated with the forgotten item satisfies:
\begin{equation*}
\left| f_{i^*}(S_u^l) - \alpha f_{\mathcal M'}(S_u^l) \right|
\le |1-\alpha| \, |f_{i^*}(S_u^l)| + \alpha \varepsilon
\end{equation*}

\noindent \textbf{Proof}:
Using the decomposition of $f_{\mathcal M}$:

\begin{equation*}
\tilde y = f_{\mathrm{base}}(S_u^l) + f_{i^*}(S_u^l)
- \alpha f_{\mathcal M'}(S_u^l).
\end{equation*}
Define the residual forgotten contribution:
\begin{equation*}
r(S_u^l) = f_{i^*}(S_u^l) - \alpha f_{\mathcal M'}(S_u^l)
\end{equation*}
Adding and subtracting $\alpha f_{i^*}(S_u^l)$:
\begin{equation*}
r(S_u^l) = (1-\alpha) f_{i^*}(S_u^l) + \alpha \Big(f_{i^*}(S_u^l) - f_{\mathcal M'}(S_u^l) \Big).
\end{equation*}
Applying triangle inequality:
\begin{equation*}
|r(S_u^l)| \le |1-\alpha| |f_{i^*}(S_u^l)| 
+ \alpha \left| f_{i^*}(S_u^l) - f_{\mathcal M'}(S_u^l) \right|
\end{equation*}
By assumption $\left| f_{i^*}(S_u^l) - f_{\mathcal M'}(S_u^l)
\right| \le \varepsilon$ and therefore:
\begin{equation*}
|r(S_u^l)| \le |1-\alpha| |f_{i^*}(S_u^l)| + \alpha\varepsilon \, \square
\end{equation*}

\noindent \textbf{Interpretation:} When $\alpha = 1$, the bound reduces to:
\begin{equation*}
\left| f_{i^*}(S_u^l) - f_{\mathcal M'}(S_u^l) \right| \le
\varepsilon
\end{equation*}
showing that the contribution of the target item is effectively removed up to the approximation error of the auxiliary model. Hence, the unlearning effectiveness of \approach depends on how well $f_{\mathcal{M}'}(S_u^l)$ approximates the local influence of the forgotten item.

\section{Experimental Settings}
\red{In this section, we present the datasets selected for our task, along with the model backbones employed, the selected competitors and the evaluation protocol adopted.}

\subsection{Evaluation Datasets}
\red{We validate the unlearning effectiveness of our approach on commonly used real-world datasets:
\begin{itemize}
    \item \textbf{MovieLens (ML)}~\cite{harper2015movielens}: A public benchmark of user--movie interactions. We use the \textbf{ML-1M} and \textbf{ML-100K} variants.
    \item \textbf{Amazon Beauty (Beauty)}~\cite{mcauley2015image}: The ``Beauty'' subset of the Amazon Reviews dataset, widely used for product recommendation tasks.
    \item \textbf{Steam}: User reviews and game metadata collected from the Steam gaming platform.
    \item \textbf{Foursquare (FS)}~\cite{6844862}: User check-in data collected over approximately ten months. We use both the New York City (\textbf{FS-NYC}) and Tokyo (\textbf{FS-TKY}) subsets.
    \item \textbf{Yelp}~\cite{asghar2016yelp}: A business review dataset released as part of the Yelp Dataset Challenge 2014.
\end{itemize}
Additional data statistics are provided in \Cref{tab:dataset_info}. We intentionally include datasets that vary in their suitability for sequential recommendation tasks, following the observations of~\cite{klenitskiy2024does}. This allows us to verify the robustness of our approach across different data characteristics. Moreover, the majority of the selected datasets have been adopted in prior work on recommendation unlearning \cite{ultrare,chen2025pre}.}

\subsection{Metrics}
To evaluate recommendation performance, we use Normalized Discounted Cumulative Gain (NDCG) \cite{10494051}, a widely adopted metric in Information Retrieval (\up is better). To measure how well the recommendation rankings are preserved after unlearning, we adopt the Finite Rank-Biased Overlap (FRBO) metric \cite{betello2024investigating}, which quantifies the agreement between two ranked lists. Higher FRBO values indicate that the post-unlearning rankings closely match the original ones, suggesting minimal disruption to the overall recommendation quality (\up is better).
To evaluate the unlearning performance, we use UnlearnRecall and UnlearnMRR from \cite{10.1145/3705328.3748092}. UnlearnRecall measures how many items that an item asked to be unlearned still appear in the user’s Top-N recommendation list (\down is better), while UnlearnMRR evaluates how strongly a model still ``remembers'' items that should be forgotten by checking how early they appear in a user’s Top-N recommendations (\down is better).

\subsection{Backbone Models}
\approach is implemented with three SRSs models: 
\begin{itemize}
   \item \textbf{GRU4Rec} \cite{hidasi2018recurrent}: this model leverages GRUs to model temporal patterns and sequential dependencies in user-item interactions.
    \item \textbf{SASRec} \cite{kang2018selfattentive}: it employs self-attention mechanisms to identify the importance and relevance of items within a user’s interaction sequence.
    \item \textbf{BERT4Rec} \cite{sun2019bert4rec}: built upon the BERT architecture, this model captures complex relationships in user behaviour sequences using bidirectional self-attention.
\end{itemize}
These representative models are the same presented in \cite{sru}. \red{These models span three distinct architectural paradigms, that is recurrent, unidirectional attention, and bidirectional transformers~\cite{pan2026survey}. Since \approach is model-agnostic by design, extending it to more recent sequential recommenders is a natural direction for future work.}

\begin{table}[t!]
    \caption{Dataset statistics after pre-processing; users and items not having at least 5 interactions are removed. Avg. and Med. refer to the Average and Median of $\frac{\mathrm{Actions}}{\mathrm{User}}$.}
    \resizebox{\columnwidth}{!}{
      \begin{tabular}{lcccccc}
        \toprule
        Name & Users & Items & Interactions &  Density(\%) & Avg. & Med. \\ 
        \midrule
        Beauty & 1,274 & 1,076 & 7,113 & 0.5190 & 5.58 & 5\\
        ML-1M &  6,040 & 3,416 & 999,611 & 4.8450 & 165 & 96\\
        ML-100k & 943 & 1,349 & 99,287 & 7.8049 & 105 & 64 \\
        Steam & 334,536 & 13,046 & 4,212,143 & 0.0009 & 12.59 & 8 \\
        FS-TKY & 2,293 & 15,177 & 494,807 & 1.422 & 215 & 146\\
        FS-NYC & 1,083 & 9,989 & 179,468 & 1.6590 & 165 & 116\\
        Yelp & 287,116 & 148,523 & 4,392,169 & 0.0103 & 15.29 & 8 \\ 
        \bottomrule
      \end{tabular}
    }
\label{tab:dataset_info}
\end{table}

\begin{table*}[t!]
\centering
\caption{Unlearning performance comparison, all @20. \textbf{Bold} denotes the best model for a dataset, \underline{underlined} the second best. $\dagger$ indicates a statistically significant result  w.r.t. the second-best approach based on Wilcoxon test with \textbf{p}-value \textbf{< 0.05}. R stands for Recall.}
\label{tab:table}
\resizebox{\linewidth}{!}{
\begin{tabular}{ll|cccc|cccc|cccc}
\toprule
\multirow{2}{*}{\textbf{Dataset}} & \multirow{2}{*}{\textbf{Method}}
& \multicolumn{4}{c|}{\textbf{SASRec}}
& \multicolumn{4}{c|}{\textbf{GRU4Rec}}
& \multicolumn{4}{c}{\textbf{BERT4Rec}} \\
\cmidrule(lr){3-6} \cmidrule(lr){7-10} \cmidrule(lr){11-14}
&
& NDCG (\up)  & UL MRR (\down) & UL R (\down) & FRBO (\up)
& NDCG (\up) & UL MRR (\down) & UL R (\down) & FRBO (\up)
& NDCG (\up) & UL MRR (\down) & UL R (\down) & FRBO (\up) \\
\midrule

\multirow{6}{*}{Beauty}
& Retrain
& \textbf{0.6484}$^{\dagger}$ & \underline{0.0016} & \underline{0.0147} & \textbf{0.7135}$^{\dagger}$
& \textbf{0.6330}$^{\dagger}$ & \textbf{0.0022} & 0.0164 & \textbf{0.6462}$^{\dagger}$
& \underline{0.6117} & 0.0035 & \textbf{0.0191} & \textbf{0.5495}$^{\dagger}$ \\

& RecEraser
& 0.5268 & 0.0024 & 0.0182 & 0.3311
& 0.4150 & 0.0025 & \underline{0.0156} & 0.2303
& 0.3566 & \underline{0.0020} & 0.0200 & 0.2417 \\

& SRU
& 0.3908 & 0.0036 & 0.0182 & 0.2936
& 0.3690 & 0.0049 & 0.0217 & 0.2131
& 0.3153 & 0.0047 & 0.0251 & 0.2960 \\

& UltraRE
& 0.4224 & 0.0016 & 0.0147 & 0.1911
& 0.1394 & 0.0025 & 0.0156 & 0.0932
& 0.1989 & \textbf{0.0013} & \textbf{0.0122} & 0.0790 \\

& IFRU
& 0.3167 & 0.0047 & 0.0156 & 0.0956
& 0.0019 & 0.0036 & 0.0217 & 0.0087
& 0.0019 & 0.0028 & 0.0200 & 0.0065 \\

& \textbf{\approach}
& \underline{0.5640} & \textbf{0.0015}$^{\dagger}$ & \textbf{0.0130}$^{\dagger}$ & \underline{0.5128}
& \underline{0.5492} & \underline{0.0024} & \textbf{0.0148}$^{\dagger}$ & \underline{0.5669}
& \textbf{0.6251}$^{\dagger}$ & 0.0028 & \underline{0.0165} & \second{0.5003} \\

\midrule

\multirow{6}{*}{ML-1M}
& Retrain
& \underline{0.1208} & 0.0018 & 0.0076 & \underline{0.3036}
& \textbf{0.1541}$^{\dagger}$ & 0.0015 & 0.0058 & \underline{0.5251}
& \textbf{0.0805}$^{\dagger}$ & \underline{0.0013} & 0.0069 & \textbf{0.6147}$^{\dagger}$ \\

& RecEraser
& 0.0417 & 0.0020 & 0.0084 & 0.0771
& 0.0459 & 0.0013 & 0.0078 & 0.1378
& 0.0034 & 0.0015 & \textbf{0.0061} & 0.1280 \\

& SRU
& 0.0259 & 0.0016 & 0.0079 & 0.0422
& 0.0285 & 0.0016 & 0.0066 & 0.1048
& 0.0280 & 0.0014 & 0.0079 & 0.1221 \\

& UltraRE
& 0.0279 & 0.0017 & 0.0084 & 0.0407
& 0.0322 & \textbf{0.0006} & \textbf{0.0048} & 0.0977
& 0.0246 & 0.0020 & 0.0083 & 0.1031 \\

& IFRU
& 0.0020 & \underline{0.0011} & \underline{0.0063} & 0.0020
& 0.0017 & \underline{0.0008} & \underline{0.0055} & 0.0023
& 0.0024 & 0.0012 & \second{0.0064} & 0.0024 \\

& \textbf{\approach}
& \textbf{0.1595}$^{\dagger}$ & \textbf{0.0008} & \textbf{0.0058}$^{\dagger}$ & \textbf{0.6080}$^{\dagger}$
& \underline{0.1235} & 0.0011 & 0.0058 & \textbf{0.5384}$^{\dagger}$
& \underline{0.0708} & \textbf{0.0009}$^{\dagger}$ & 0.0071 & \underline{0.4821} \\

\midrule

\multirow{6}{*}{ML-100k}
& Retrain
& \textbf{0.0945}$^{\dagger}$ & 0.0043 & \underline{0.0156} & \underline{0.6017}
& \textbf{0.0991} & 0.0043 & 0.0156 & \textbf{0.7514}$^{\dagger}$
& \textbf{0.0658} & \textbf{0.0018}$^{\dagger}$ & 0.0123 & \textbf{0.5211}$^{\dagger}$ \\

& RecEraser
& 0.0452 & 0.0034 & 0.0190 & 0.1287
& 0.0422 & 0.0058 & 0.0212 & 0.1264
& 0.0328 & \underline{0.0010} & 0.0122 & 0.1065 \\

& SRU
& 0.0490 & \underline{0.0031} & 0.0201 & 0.1360
& 0.0504 & 0.0046 & 0.0178 & 0.1641
& 0.0464 & 0.0032 & 0.0178 & 0.0924 \\

& UltraRE
& 0.0471 & \textbf{0.0028} & 0.0201 & 0.1202
& 0.0597 & 0.0040 & \underline{0.0156} & 0.1802
& 0.0362 & 0.0013 & \underline{0.0100} & 0.1188 \\

& IFRU
& 0.0053 & 0.0058 & 0.0188 & 0.0060
& 0.0095 & \underline{0.0035} & 0.0178 & 0.0049
& 0.0036 & 0.0031 & 0.0189 & 0.0055 \\

& \textbf{\approach}
& \underline{0.0864} & 0.0041 & \textbf{0.0145}$^{\dagger}$ & \textbf{0.6653}$^{\dagger}$
& \underline{0.0871} & \textbf{0.0027}$^{\dagger}$ & \textbf{0.0156} & \underline{0.5580}
& \underline{0.0490} & 0.0023 & \textbf{0.0100} & \underline{0.4053} \\

\midrule
\multirow{6}{*}{Steam}
& Retrain
& \textbf{0.0832}$^{\dagger}$ & \textbf{0.0003} & \textbf{0.0017} & \textbf{0.4385}$^{\dagger}$
& \textbf{0.0424}$^{\dagger}$ & \underline{0.0003} & \underline{0.0018} & \textbf{0.4468}$^{\dagger}$
& \textbf{0.0793}$^{\dagger}$ & \textbf{0.0003} & 0.0019 & \textbf{0.6488}$^{\dagger}$ \\

& RecEraser
& 0.0537 & 0.0007 & 0.0020 & 0.0655
& 0.0333 & 0.0005 & 0.0023 & 0.0670
& 0.0293 & 0.0005 & 0.0022 & 0.0917 \\

& SRU
& 0.0571 & 0.0005 & 0.0022 & 0.0878
& \underline{0.0409} & 0.0004 & 0.0022 & 0.1137
& 0.0421 & 0.0006 & 0.0023 & 0.1946 \\

& UltraRE
& 0.0614 & 0.0004 & 0.0019 & 0.0594
& 0.0392 & 0.0004 & 0.0021 & 0.0674
& 0.0325 & 0.0004 & 0.0024 & 0.1445 \\

& IFRU
& 0.0669 & 0.0006 & \underline{0.0018} & 0.0188
& 0.0008 & 0.0005 & 0.0020 & 0.0006
& 0.0008 & 0.0004 & 0.0022 & 0.0004 \\

& \textbf{\approach}
& \underline{0.0721} & \underline{0.0004} & 0.0020 & \underline{0.3319}
& 0.0189 & \textbf{0.0002} & \textbf{0.0016} & \underline{0.1645}
& \underline{0.0423} & \underline{0.0003} & \underline{0.0020} & \underline{0.2542} \\

\midrule
\multirow{6}{*}{FS-TKY}
& Retrain
& \underline{0.4199} & 0.0011 & 0.0051 & \textbf{0.6566}$^{\dagger}$
& \textbf{0.3161}$^{\dagger}$ & 0.0007 & \textbf{0.0014}$^{\dagger}$ & \textbf{0.7801}$^{\dagger}$
& \textbf{0.2773}$^{\dagger}$ & 0.0006 & 0.0022 & \textbf{0.6780}$^{\dagger}$ \\

& RecEraser
& 0.2452 & \underline{0.0009} & 0.0028 & 0.1978
& 0.1591 & 0.0008 & 0.0028 & 0.1804
& \underline{0.1654} & 0.0004 & 0.0027 & \underline{0.2782} \\

& SRU
& 0.0912 & 0.0011 & \underline{0.0023} & 0.2123
& 0.0908 & 0.0015 & 0.0023 & 0.2563
& 0.0905 & 0.0006 & 0.0032 & 0.2638 \\

& UltraRE
& 0.1420 & 0.0010 & 0.0036 & 0.1839
& 0.1025 & 0.0009 & 0.0028 & 0.2193
& 0.1223 & 0.0005 & 0.0022 & 0.1955 \\

& IFRU
& 0.1253 & \textbf{0.0002}$^{\dagger}$ & \textbf{0.0014}$^{\dagger}$ & 0.0341
& 0.0008 & \underline{0.0002} & 0.0019 & 0.0008
& 0.0001 & \underline{0.0001} & \textbf{0.0004}$^{\dagger}$ & 0.0006 \\

& \textbf{\approach}
& \textbf{0.4371} & 0.0012 & 0.0028 & \underline{0.5972}
& \underline{0.2487} & \textbf{0.0002} & \underline{0.0018} & \underline{0.2598}
& 0.1238 & \textbf{0.0001} & \underline{0.0014} & 0.1252 \\

\midrule
\multirow{6}{*}{FS-NYC}
& Retrain
& \underline{0.3565} & 0.0011 & 0.0029 & \textbf{0.6013}$^{\dagger}$
& \textbf{0.2892} & 0.0002 & 0.0019 & \textbf{0.6917}$^{\dagger}$
& \textbf{0.2115}$^{\dagger}$ & 0.0003 & 0.0048 & \textbf{0.6132}$^{\dagger}$ \\

& RecEraser
& 0.1473 & 0.0017 & 0.0039 & 0.1612
& 0.0979 & 0.0003 & 0.0020 & 0.1109
& 0.0803 & 0.0003 & 0.0009 & 0.1593 \\

& SRU
& 0.0396 & \underline{0.0001} & 0.0010 & 0.1424
& 0.0375 & \textbf{0.0001} & \textbf{0.0010} & 0.1780
& 0.0368 & \underline{0.0002} & \underline{0.0019} & 0.1698 \\

& UltraRE
& 0.1057 & \textbf{0.0001} & 0.0010 & 0.1002
& 0.0904 & 0.0002 & 0.0010 & 0.0902
& 0.0533 & 0.0009 & 0.0048 & 0.0742 \\

& IFRU
& 0.1218 & 0.00022 & \underline{0.0018} & 0.0322
& 0.0008 & \underline{0.0001} & \underline{0.0010} & 0.0006
& 0.0008 & \textbf{0.0002} & \textbf{0.0020} & 0.0004 \\

& \textbf{\approach}
& \textbf{0.3780}$^{\dagger}$ & 0.0003 & \textbf{0.0019} & \underline{0.5004}
& \underline{0.2575} & 0.0005 & 0.0029 & \underline{0.3192}
& \underline{0.1773} & 0.0004 & 0.0029 & \underline{0.3321} \\

\midrule
\multirow{6}{*}{Yelp}
& Retrain
& \textbf{0.0337} & 0.0003 & 0.0006 & \underline{0.4560}
& \textbf{0.0227}$^{\dagger}$ & 0.0003 & 0.0004 & \textbf{0.3860}$^{\dagger}$
& \textbf{0.0188}$^{\dagger}$ & 0.0001 & \underline{0.0004} & \textbf{0.5503} \\

& RecEraser
& 0.0132 & 0.0002 & 0.0004 & 0.0458
& 0.0103 & \textbf{0.0001} & \underline{0.0003} & 0.0421
& 0.0063 & \underline{0.0001} & 0.0006 & 0.0165 \\

& SRU
& 0.0075 & \underline{0.0001} & \textbf{0.0003} & 0.0175
& 0.0046 & \underline{0.0001} & \textbf{0.0003} & 0.0187
& 0.0031 & 0.0002 & 0.0005 & 0.0080 \\

& UltraRE
& 0.0077 & 0.0001 & 0.0004 & 0.0157
& 0.0051 & 0.0002 & 0.0005 & 0.0098
& 0.0026 & 0.0002 & 0.0005 & 0.0089 \\

& IFRU
& 0.0121 & 0.0002 & 0.0005 & 0.0041
& 0.0001 & 0.0003 & 0.0006 & 0.0001
& 0.0003 & 0.0002 & 0.0006 & 0.0001 \\

& \textbf{\approach}
& \underline{0.0304} & \textbf{0.0001} & \underline{0.0004} & \textbf{0.4867}$^{\dagger}$
& \underline{0.0145} & 0.0002 & 0.0004 & \underline{0.2980}
& \underline{0.0122} & \textbf{0.0001} & \textbf{0.0004} & \underline{0.2513} \\

\bottomrule
\end{tabular}}
\end{table*}

\subsection{Baseline Methods}

We compare \approach with \red{the} state-of-the-art baselines presented in \Cref{sec:related_unlearning}. We were unable to implement all of the methods because some did not provide access to their codebases. To enable unlearning, every baseline is trained with the backbone models defined in the previous section:

\begin{itemize}
    \item \textbf{Retrain} denotes full model retraining from scratch.
    \item \textbf{RecEraser} \cite{receraser} partitions the training data into shards and aggregates models trained on each shard.
    \item \textbf{UltraRE} \cite{ultrare}, similar to RecEraser, incorporates transport weights into the clustering process.
    \item \textbf{SRU} \cite{sru} is designed for session-based recommendation; it shards training data based on embedding similarity and aggregates the resulting models.
    \item \textbf{IFRU} \cite{ifru} is an influence-based method that performs unlearning via a dedicated unlearning loss.
\end{itemize}

\red{A natural question is whether simply removing the target item from the output list, commonly referred to as post-hoc filtering, constitutes a valid unlearning strategy.} Post-hoc filtering operates at the output level by excluding the target item from the recommendation list, leaving the model's internal representations unchanged.
To verify this, let $g_{\mathcal{M}}: \mathcal{I}^l \rightarrow \mathbb{R}^d$ denote the function mapping an input sequence $S_u^l$ to its last hidden state before the output layer. We define $h = g_{\mathcal{M}}(S_u^l)$ and $h^{\text{U}} = g_{\mathcal{M}^{U}}(S_u^l)$ as the hidden states of the original and unlearned models, respectively, and measure $||h - h^{\text{U}}||_2$ to quantify how each method modifies internal representations.
Post-hoc filtering yields a distance of zero across all sequences, confirming that the forgotten item's influence remains encoded in the hidden representations. In contrast, \approach produces non-zero distances, demonstrating that it operates at the representation level rather than as a post-processing step. Hence, we exclude post-hoc filtering as a competitor.

\subsection{Implementation Settings}
We used EasyRec library\cite{betello2024reproducible} to perform our experiments. Our pre-processing strategy follows established practices, such as treating ratings as implicit and removing users and items with fewer than 5 interactions \cite{kang2018selfattentive, sun2019bert4rec}.
For testing, as in \cite{sun2019bert4rec, kang2018selfattentive}, we keep the last interaction for each user, while for the validation set, the second to last action is retained. All remaining interactions contribute to the training set.
Each observed user-item interaction is treated as a positive instance, and negative items are sampled from items not previously consumed by the user \cite{krichene2020sampled}. We set the sequence length to 200 for the MovieLens and Foursquare datasets, and to 50 for the Steam, Amazon, and Yelp datasets, following the results of previous work~\cite{10.1145/3631116}.
We evaluate unlearning on five different target items and report the average results across them. \red{We set the batch size to 128 and optimize all models using the Adam optimizer. For each method, we perform a grid search to select the best hyperparameters. All models are trained for 500 epochs, while methods requiring model aggregation are additionally trained for 10 aggregation epochs, as in the original papers. After a thorough empirical analysis, we set $\alpha = 0.7$ and $k = 1$. Experiments are conducted on a machine equipped with an Intel Xeon W-2245 CPU (3.90GHz) and an NVIDIA RTX A6000 GPU with 48GB of VRAM.}

\section{Results}

\begin{table}[t]
\centering
\caption{Relative training time (TT) and inference time (IT) w.r.t. Retrain. \textbf{Bold} denotes the best model for a dataset, \underline{underlined} the second best. Lower is better.}
\label{tab:inference_time}
\resizebox{\linewidth}{!}{
\begin{tabular}{l|l|cc|cc|cc}
\toprule
\multirow{2}{*}{\textbf{Dataset}}
& \multirow{2}{*}{\textbf{Method}}
& \multicolumn{2}{c|}{\textbf{SASRec}}
& \multicolumn{2}{c|}{\textbf{GRU4Rec}}
& \multicolumn{2}{c}{\textbf{BERT4Rec}} \\
\cmidrule(lr){3-4}
\cmidrule(lr){5-6}
\cmidrule(lr){7-8}
& & TT ($\downarrow$) & IT ($\downarrow$)
& TT ($\downarrow$) & IT ($\downarrow$)
& TT ($\downarrow$) & IT ($\downarrow$) \\
\midrule

\multirow{6}{*}{Beauty}
& Retrain   & 1.00$\times$ & 1.00$\times$ & 1.00$\times$ & 1.00$\times$ & 1.00$\times$ & 1.00$\times$ \\
& RecEraser & 3.38$\times$ & 1.21$\times$ & 3.52$\times$ & \underline{1.09$\times$} & 2.77$\times$ & \underline{1.06$\times$} \\
& SRU       & 6.90$\times$ & \underline{1.18$\times$} & 5.72$\times$ & 1.14$\times$ & 54.49$\times$ & 1.07$\times$ \\
& UltraRE   & 8.70$\times$ & 1.65$\times$ & 13.28$\times$ & 1.32$\times$ & 13.29$\times$ & 1.36$\times$ \\
& IFRU      & \textbf{0.10$\times$} & 1.66$\times$ & \textbf{0.07$\times$} & 1.67$\times$ & \textbf{0.12$\times$} & 1.59$\times$ \\
& \approach & \underline{0.89$\times$} & \textbf{1.11$\times$} & \underline{0.90$\times$} & \textbf{1.08$\times$} & \underline{0.84$\times$} & \textbf{1.05$\times$} \\
\midrule

\multirow{6}{*}{ML1M}
& Retrain   & 1.00$\times$ & 1.00$\times$ & 1.00$\times$ & 1.00$\times$ & 1.00$\times$ & 1.00$\times$ \\
& RecEraser & 10.21$\times$ & 1.29$\times$ & 11.32$\times$ & 1.12$\times$ & 7.82$\times$ & \underline{1.63$\times$} \\
& SRU       & 12.47$\times$ & 1.16$\times$ & 12.79$\times$ & \underline{1.10$\times$} & 9.66$\times$ & 1.45$\times$ \\
& UltraRE   & 17.99$\times$ & 1.39$\times$ & 16.22$\times$ & 1.14$\times$ & 13.40$\times$ & 1.69$\times$ \\
& IFRU      & \textbf{0.06$\times$} & 1.69$\times$ & \textbf{0.07$\times$} & 1.77$\times$ & \textbf{0.10$\times$} & 1.80$\times$ \\
& \approach & \underline{0.50$\times$} & \textbf{1.05$\times$} & \underline{0.66$\times$} & \textbf{1.05$\times$} & \underline{0.51$\times$} & \textbf{1.10$\times$} \\
\midrule

\multirow{6}{*}{ML-100K}
& Retrain   & 1.00$\times$ & 1.00$\times$ & 1.00$\times$ & 1.00$\times$ & 1.00$\times$ & 1.00$\times$ \\
& RecEraser & 3.64$\times$ & 1.18$\times$ & 4.18$\times$ & 1.11$\times$ & 3.98$\times$ & 1.19$\times$ \\
& SRU       & 2.63$\times$ & \underline{1.07$\times$} & 1.70$\times$ & \underline{1.06$\times$} & 2.91$\times$ & \textbf{1.03$\times$} \\
& UltraRE   & 4.23$\times$ & 1.55$\times$ & 4.83$\times$ & 1.98$\times$ & 4.71$\times$ & 1.42$\times$ \\
& IFRU      & \textbf{0.45$\times$} & 1.64$\times$ & \textbf{0.42$\times$} & 1.70$\times$ & \textbf{0.46$\times$} & 1.77$\times$ \\
& \approach & \underline{0.58$\times$} & \textbf{1.06$\times$} & \underline{0.69$\times$} & \textbf{1.05$\times$} & \underline{0.65$\times$} & \underline{1.15$\times$} \\
\midrule

\multirow{6}{*}{Steam}
& Retrain   & 1.00$\times$ & 1.00$\times$ & 1.00$\times$ & 1.00$\times$ & 1.00$\times$ & 1.00$\times$ \\
& RecEraser & 2.28$\times$ & 1.09$\times$ & 2.25$\times$ & 1.40$\times$ & 2.28$\times$ & 1.57$\times$ \\
& SRU       & 1.34$\times$ & \textbf{0.96$\times$} & 1.31$\times$ & 1.02$\times$ & 1.35$\times$ & 1.03$\times$ \\
& UltraRE   & 2.24$\times$ & 1.11$\times$ & 2.39$\times$ & 1.41$\times$ & 6.81$\times$ & 1.61$\times$ \\
& IFRU      & \textbf{0.02$\times$} & 2.04$\times$ & \textbf{0.02$\times$} & 1.98$\times$ & \textbf{0.01$\times$} & 2.02$\times$ \\
& \approach & \underline{0.04$\times$} & \underline{1.02$\times$} & \underline{0.04$\times$} & 0.99$\times$ & \underline{0.04$\times$} & \textbf{1.01$\times$} \\
\midrule

\multirow{6}{*}{FS-TKY}
& Retrain   & 1.00$\times$ & 1.00$\times$ & 1.00$\times$ & 1.00$\times$ & 1.00$\times$ & 1.00$\times$ \\
& RecEraser & 3.13$\times$ & \underline{1.05$\times$} & 3.00$\times$ & 1.05$\times$ & 3.14$\times$ & \underline{1.04$\times$} \\
& SRU       & 5.72$\times$ & 1.07$\times$ & 5.17$\times$ & \underline{1.02$\times$} & 5.74$\times$ & 1.06$\times$ \\
& UltraRE   & 3.31$\times$ & 1.27$\times$ & 3.11$\times$ & 1.10$\times$ & 3.29$\times$ & 1.16$\times$ \\
& IFRU      & \textbf{0.11$\times$} & 2.04$\times$ & \textbf{0.09$\times$} & 2.04$\times$ & \textbf{0.11$\times$} & 2.03$\times$ \\
& \approach & \underline{0.36$\times$} & \textbf{1.04$\times$} & \underline{0.35$\times$} & \textbf{1.00$\times$} & \underline{0.36$\times$} & \textbf{1.01$\times$} \\
\midrule

\multirow{6}{*}{FS-NYC}
& Retrain   & 1.00$\times$ & 1.00$\times$ & 1.00$\times$ & 1.00$\times$ & 1.00$\times$ & 1.00$\times$ \\
& RecEraser & 2.87$\times$ & 1.06$\times$ & 2.98$\times$ & 1.05$\times$ & 2.91$\times$ & \underline{1.04$\times$} \\
& SRU       & 3.80$\times$ & \textbf{1.00$\times$} & 3.73$\times$ & \textbf{0.97$\times$} & 2.68$\times$ & \textbf{0.98$\times$} \\
& UltraRE   & 4.75$\times$ & 1.40$\times$ & 4.75$\times$ & 1.14$\times$ & 4.78$\times$ & 1.29$\times$ \\
& IFRU      & \textbf{0.25$\times$} & 1.86$\times$ & \textbf{0.22$\times$} & 1.88$\times$ & \textbf{0.26$\times$} & 1.87$\times$ \\
& \approach & \underline{0.57$\times$} & \underline{1.09$\times$} & \underline{0.60$\times$} & \underline{1.04$\times$} & \underline{0.59$\times$} & 1.07$\times$ \\
\midrule

\multirow{6}{*}{YELP}
& Retrain   & 1.00$\times$ & 1.00$\times$ & 1.00$\times$ & 1.00$\times$ & 1.00$\times$ & 1.00$\times$ \\
& RecEraser & 2.22$\times$ & 3.67$\times$ & 2.17$\times$ & 4.13$\times$ & 2.20$\times$ & 2.90$\times$ \\
& SRU       & 1.30$\times$ & 3.41$\times$ & 1.27$\times$ & 3.18$\times$ & 1.30$\times$ & \underline{1.04$\times$} \\
& UltraRE   & \underline{1.18$\times$} & \underline{1.56$\times$} & \underline{1.15$\times$} & \underline{1.71$\times$} & \underline{1.17$\times$} & 1.22$\times$ \\
& IFRU      & \textbf{0.01$\times$} & 4.89$\times$ & \textbf{0.01$\times$} & 4.22$\times$ & \textbf{0.01$\times$} & 2.01$\times$ \\
& \approach & \underline{0.04$\times$} & \textbf{0.11$\times$} & \underline{0.04$\times$} & \textbf{0.11$\times$} & \underline{0.04$\times$} & \textbf{0.98$\times$} \\

\bottomrule
\end{tabular}
}
\end{table}

In our work we aim to answer the following research questions:
\noindent\textbf{RQ1.} \red{How competitive is \approach, based on influence subtraction on item unlearning in SRSs, w.r.t. SOTA models in terms of unlearning effectiveness and recommendation quality preservation?}

\noindent\textbf{RQ2.} How efficient is \approach, \red{in terms of training and inference time} and how does it compare with its competitors?

\noindent\textbf{RQ3.} What's the impact of hyperparameters \red{$\alpha$ and $k$} on performance?

\noindent\textbf{RQ4.} To what extent are the theoretical properties of \approach empirically observed in practice?

\subsection*{RQ1: Effectiveness of \approach}
We first assess whether \approach\ can effectively remove the influence of target entities while preserving recommendation quality.  

\begin{figure}[t]
    \centering
    \includegraphics[width=\columnwidth]{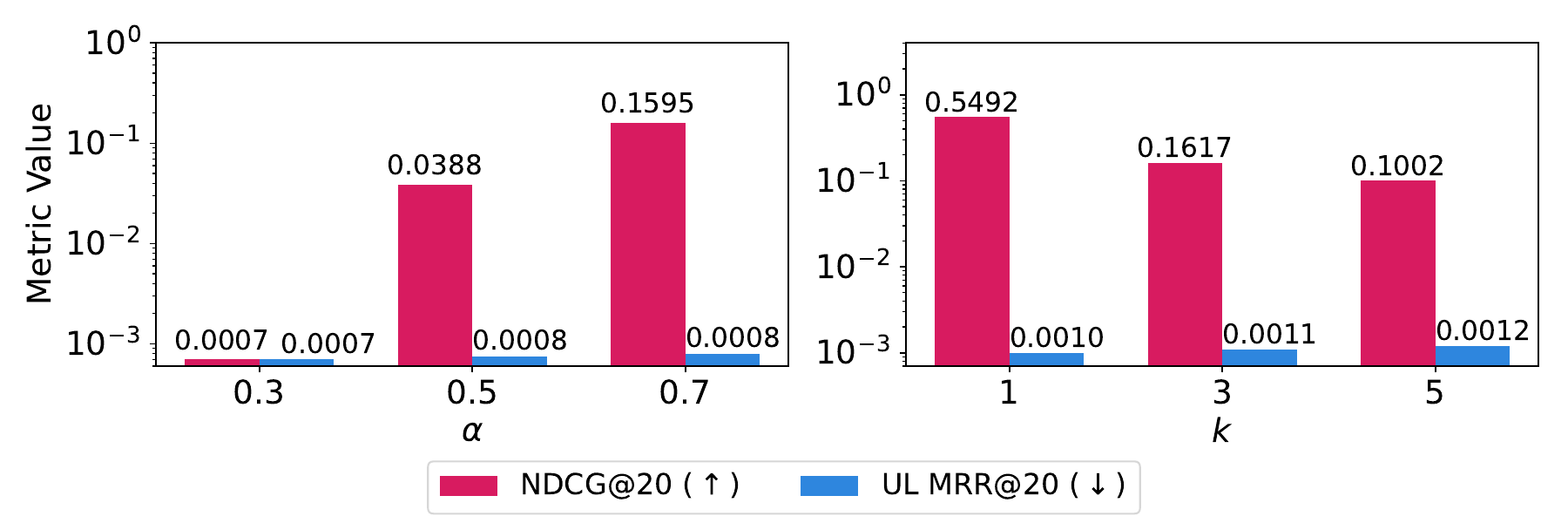}
    \caption{Effect of the hyperparameters $k$ (left) and $\alpha$ (right) on recommendation (red) and unlearning (blue) performance.}
\label{fig:ablation}
\end{figure}

\begin{figure}[t]
    \centering
    \includegraphics[width=\columnwidth]{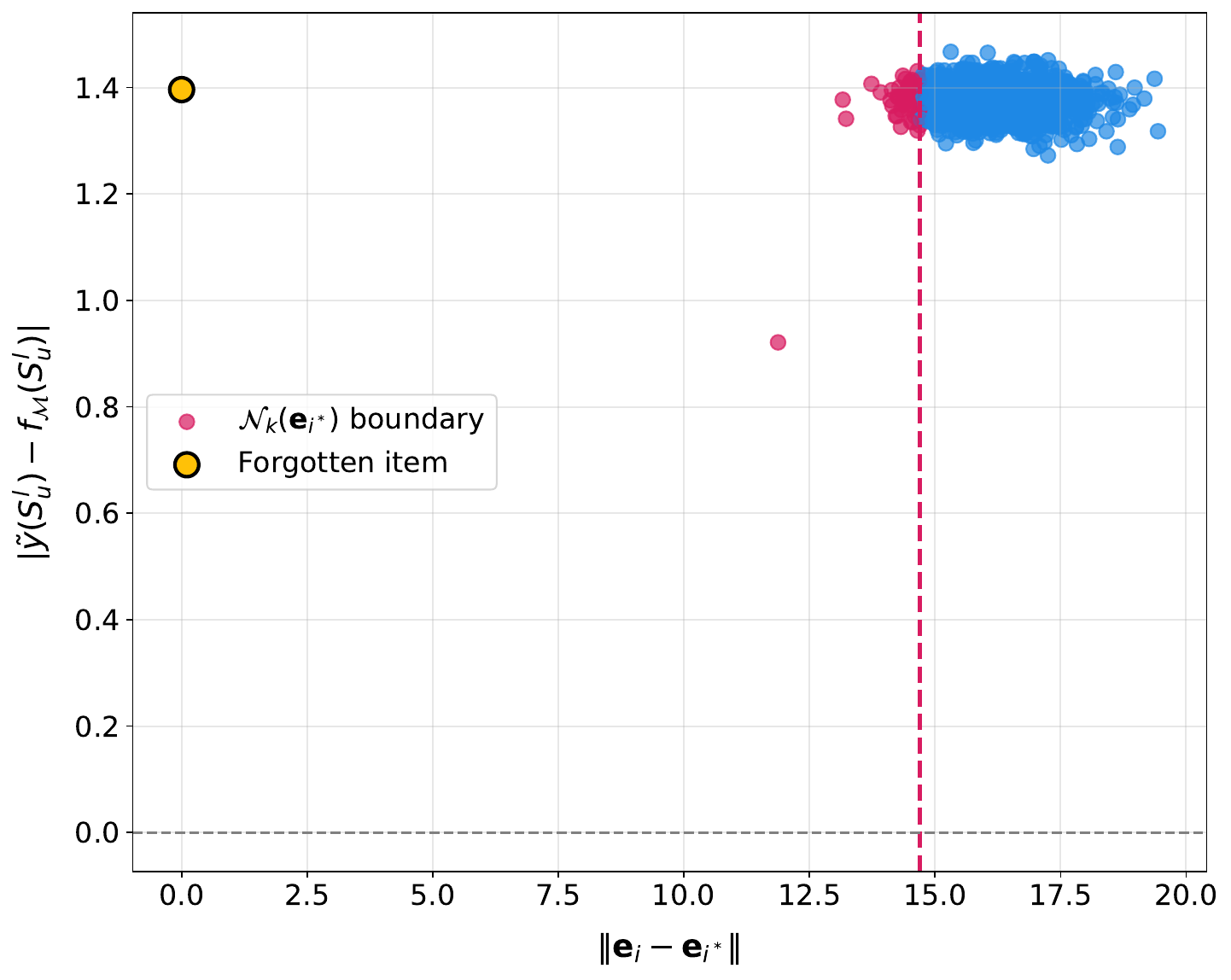}
    \caption{Average score shift magnitude as a function of the embedding distance from the forgotten item. The orange dashed line denotes the boundary of the neighborhood $\mathcal{N}_k(\mathbf{e}_{i^*})$. Results are obtained on the ML-100k dataset with the SASRec backbone.}
    \label{fig:theorem2}
\end{figure}

\begin{figure*}[t]
    \centering
    \includegraphics[width=\linewidth]{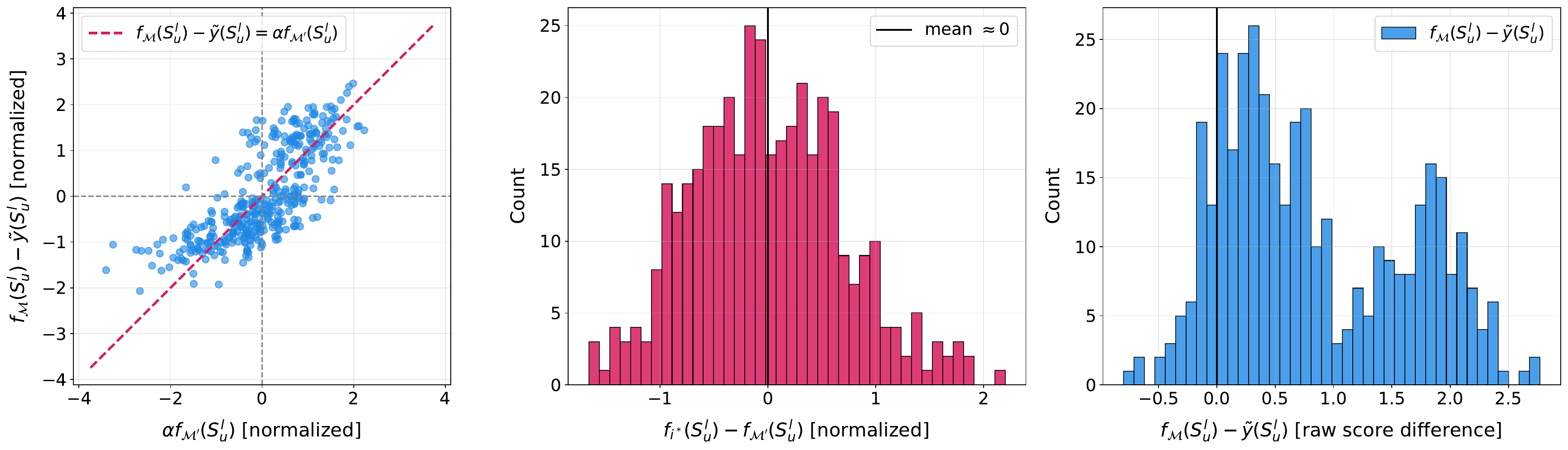}
    \caption{Empirical validation of influence removal. \textbf{Left:} Scatter plot comparing the score difference $f_{\mathcal{M}}(S_u^l) - \tilde{y}(S_u^l)$ with the scaled auxiliary model output $\alpha f_{\mathcal{M}'}(S_u^l)$. \textbf{Center:} Distribution of the residual $f_{i^*}(S_u^l) - f_{\mathcal{M}'}(S_u^l)$, showing a tight concentration around zero (mean $\approx 0$). \textbf{Right:} Distribution of the raw score difference $f_{\mathcal{M}}(S_u^l) - \tilde{y}(S_u^l)$, with a positive mean.}
    \label{fig:theorem3}
\end{figure*}

In \Cref{tab:table} we can see the results: across all backbones and datasets, \approach consistently achieves the best or second-best unlearning metrics (UL-MRR and UL-Recall).  
On SASRec, \approach\ attains the lowest unlearning scores on Beauty and ML-1M, even outperforming full retraining in several cases, indicating a more complete removal of the target signal. Similar trends hold for GRU4Rec and BERT4Rec, where \approach\ maintains competitive or superior forgetting compared to baselines such as RecEraser, SRU, and UltraRE.

Importantly, this strong unlearning does not come at the expense of utility. \approach preserves high recommendation quality, achieving near-retrain NDCG on Beauty and often surpassing retraining on ML-1M (e.g., with SASRec and BERT4Rec). FRBO further confirms that \approach maintains ranking stability relative to retraining while removing the target influence. These results show that \approach strikes a balance between effective forgetting and utility preservation.

\subsection*{RQ2: Efficiency of \approach}
We next compare \approach against retraining and prior unlearning methods in terms of both training and inference efficiency. Compared to retraining, \approach achieves comparable or better unlearning quality while substantially reducing computational cost, as shown in \cref{tab:inference_time}. Whereas retraining requires full model optimization, \approach reduces training time by up to $\sim$2$\times$ on ML-1M and remains consistently faster across datasets. Methods such as RecEraser and SRU are generally slower while also underperforming in either unlearning effectiveness or recommendation quality.
Although IFRU is the most computationally efficient approach during training, it severely degrades recommendation performance, often causing NDCG to collapse. In contrast, \approach offers a more favorable trade-off by preserving recommendation accuracy, achieving state-of-the-art unlearning performance, and maintaining high computational efficiency.
Beyond training efficiency, \cref{tab:inference_time} shows that \approach also consistently achieves the lowest inference time among all compared methods. \approach achieves the lowest inference time in the majority of settings, with the gap widening on larger datasets. On Yelp, inference ratios for competing methods range from $1.22\times$ to $4.89\times$, whereas \approach requires only $0.11\times$ for SASRec and GRU4Rec, nearly an order of magnitude faster. Notably, despite its fast training, IFRU consistently inflates inference time (up to $4.89\times$), suggesting its approximate updates do not yield inference-friendly representations.
Overall, these findings indicate that \approach enables scalable and reliable unlearning by jointly minimizing training and inference costs, while preserving strong recommendation quality and effective forgetting behavior.

\subsection*{RQ3: Hyperparameters Impact}

\Cref{fig:ablation} shows the impact of the two main hyperparameters: \red{$k$ (number of nearest neighbors of the item to be forgotten) and $\alpha$ (scaling hyperparameter controlling the unlearning strength)}. From the left plot, increasing $k$ significantly degrades recommendation quality, while unlearning effectiveness remains relatively stable. A smaller $k$ preserves more of the original model's utility by limiting the influence subtraction to a tighter neighborhood around the forgotten item. In our experiments, we select $k=1$ as it achieves the best trade-off, maintaining high NDCG while ensuring effective unlearning. Higher values of $k$ may be more suitable in scenarios where the item to be forgotten is relatively unpopular.

From the right plot, lower $\alpha$ values aggressively suppress recommendations, yielding near-zero NDCG but strong unlearning. Higher values preserve recommendation quality, with a marginal increase in UL MRR@20. This confirms that $\alpha$ modulates the trade-off between utility preservation and forgetting strength.

\subsection*{RQ4: Empirical Validation of Theoretical Guarantees}

\paragraph{\textbf{Ranking Stability}}

The theoretical result in~\cref{sec:stability} suggests that sufficiently large
score margins preserve the relative ranking after unlearning.
To empirically assess this behavior, we analyze the FRBO~\cite{betello2024investigating}.
As shown in \cref{tab:table}, \approach consistently achieves high FRBO scores across datasets and backbone models, suggesting that the relative ordering of relevant items remains largely unchanged after unlearning. These results indicate that the user--item relationships learned by the original model are mostly preserved, limiting unnecessary distortions in recommendation quality. Overall, the findings support that the proposed method effectively removes the targeted information while maintaining ranking stability.

\begin{tcolorbox}[
    colback=green!5,
    colframe=green!30,
    left=1mm
]
\begin{tabular}{@{}m{0.02\linewidth}m{0.90\linewidth}@{}}
\centering \emoji{check-mark-button} &
\approach preserves the relative ordering among relevant items and attenuates the influence of forgotten entities.
\end{tabular}
\end{tcolorbox}

\paragraph{\textbf{Local Perturbation}}
To evaluate the practical implications of the theoretical results on local perturbation, we measure the average score shift magnitude $|\tilde{y}(S_u^l) - f_{\mathcal{M}}(S_u^l)|$ for samples whose embeddings lie at different distances from the forgotten item embedding $e_{i^*}$. For each sample, the x-axis reports the embedding distance $\|e_u - e_{i^*}\|$ while the y-axis reports the induced score perturbation after applying the unlearning update $\tilde{y}(S_u^l) = f_{\mathcal{M}}(S_u^l) - \alpha f_{\mathcal{M}'}(S_u^l)$. Results are obtained on the ML-100k dataset with the SASRec backbone.

\Cref{fig:theorem2} shows that, within the neighborhood $\mathcal{N}_k(\mathbf{e}_{i^*})$, the modification introduced by unlearning remains limited and approximately constant. In particular, no sharp increase in score perturbation is observed near the boundary of the neighborhood (orange dashed line), suggesting that the update behaves smoothly around the forgotten item. This empirical behavior is consistent with the theory, which predicts an upper bound proportional to the neighborhood diameter $\alpha\left(L\,\mathrm{diam}(\mathcal{N}_k(\mathbf{e}_{i^*})) + C\right)$.
Although the theorem provides only a worst-case guarantee, the observed perturbations remain well below this bound, indicating that the practical effect of unlearning is more stable than the theoretical maximum.

\begin{tcolorbox}[
    colback=green!5,
    colframe=green!30,
    left=1mm
]
\begin{tabular}{@{}m{0.02\linewidth}m{0.90\linewidth}@{}}
\centering \emoji{check-mark-button} &
\approach induces only bounded local modifications of the scoring function around the forgotten item.
\end{tabular}
\end{tcolorbox}

\paragraph{\textbf{Influence Removal}}
\Cref{fig:theorem3} empirically validates the theoretical bound on influence removal. The left scatter plot compares the normalized score difference $f_{\mathcal{M}}(S_u^l) - \tilde{y}(S_u^l)$ with the scaled auxiliary model output $\alpha f_{\mathcal{M}'}(S_u^l)$. We observe that the points lie close to the diagonal, indicating that the auxiliary model accurately approximates the contribution of the target item, i.e., $f_{\mathcal{M}'}(S_u^l) \approx f_{i^*}(S_u^l)$. This directly supports the approximation assumption underlying the theory.

The center histogram shows the distribution of the residual $f_{i^*}(S_u^l) - f_{\mathcal{M}'}(S_u^l)$. The residuals are tightly concentrated around zero (mean $\approx 0$), demonstrating that the approximation error $\varepsilon$ is small in practice. This confirms that the bound $\left| f_{i^*}(S_u^l) - f_{\mathcal{M}'}(S_u^l) \right| \leq \varepsilon$ is not only theoretically justified but also empirically tight.

Finally, the right histogram reports the distribution of $f_{\mathcal{M}}(S_u^l) - \tilde{y}(S_u^l)$, with a positive mean (approximately $0.83$), indicating that the removed component has a substantial contribution to the original predictions. This shows that the unlearning operation is both non-trivial and effective.
In particular, when $\alpha = 1$, the empirical residual remains small, confirming that the contribution of the target item is effectively removed up to a negligible approximation error.

\begin{tcolorbox}[
    colback=green!5,
    colframe=green!30,
    left=1mm
]
\begin{tabular}{@{}m{0.02\linewidth}m{0.90\linewidth}@{}}
\centering \emoji{check-mark-button} &
Since the auxiliary model learns almost exactly the same signal as the target item, subtracting it cancels that signal out, leaving only a tiny leftover error.
\end{tabular}
\end{tcolorbox}


\section{Conclusions}
In this paper, we introduced \approach, a lightweight and model-agnostic framework for machine unlearning in Sequential Recommender Systems. Unlike existing approaches based on costly retraining or influence-function approximations, \approach removes the contribution of forgotten entities through a subtractive mechanism: it identifies the local embedding neighborhood of the target item, trains a compact auxiliary model on this restricted region, and subtracts its influence during inference. \red{Moreover, from a practical perspective, \approach is particularly attractive in dynamic production environments, where unlearning requests may later need to be reverted. Because the original recommender parameters remain untouched, forgotten items can be recovered instantly without retraining costs.}

Extensive experiments on multiple real-world datasets and sequential recommendation architectures demonstrate that \approach matches or surpasses existing baselines in unlearning metrics while preserving ranking quality and requiring only a fraction of the retraining cost. Notably, \approach achieves up to 32\% improvement in NDCG over the best-performing baseline and attains FRBO values similar or better than those of full retraining on several datasets, confirming strong ranking preservation. At the same time, \approach requires \red{just 2\% of the original retraining baseline time budget}, while shard-based methods incur up to 17$\times$ overhead.

As future work, we plan to extend \approach into other domains such as image and text.

\begin{acks}
We would like to thank Maria Diana Calugaru for helpful discussions in early stages of this project. This work was partially supported by the project SEEDS: Sustainable Ecosystems in Evolving Digital Societies PROGETTI DIPARTIMENTALI DI ATENEO. This work was partially supported by the FoReLab and CrossLab projects (Departments of Excellence), and the NVIDIA Academic Grants Program 2026.
\end{acks}

\balance
\bibliographystyle{ACM-Reference-Format}
\bibliography{biblio}

\end{document}